\documentstyle[11pt,graphicx]{article}
\title{Alleviation of Cosmic Age Problem
         In Interacting Dark Energy Model}
\author{S. Wang\footnote{Email: swang@mail.ustc.edu.cn}
        \,and Y. Zhang\\
        Astrophysics Center \\
        University of Science and Technology of China \\
        Hefei, Anhui, China }
 \date{}

\begin{document}
\maketitle
\baselineskip=19truept

\def\vek{\vec{k}}
\renewcommand{\arraystretch}{1.5}
\newcommand{\be}{\begin{equation}}
\newcommand{\ee}{\end{equation}}
\newcommand{\ba}{\begin{eqnarray}}
\newcommand{\ea}{\end{eqnarray}}


\sf

\begin{center}
\Large  Abstract
\end{center}

\baselineskip=19truept

\begin{quote}
{\large We investigate the cosmic age problem
associated with the old high-$z$ quasar APM 08279 + 5255
and the oldest globular cluster M 107,
both being difficult to accommodate in $\Lambda$CDM model.
By evaluating the age of the Universe
in a model that has an extremely phantom like form of dark energy (DE),
we show that simply introducing the dark energy alone does not remove the problem,
and the interaction between dark matter (DM) and DE need to be taken into account.
Next, as examples, we consider two interacting DE models.
It is found that both these two interacting DE Models
can predict a cosmic age much greater than that of $\Lambda$CDM model at any redshift,
and thus substantially alleviate the cosmic age problem.
Therefore, the interaction between DM and DE
is the crucial factor required
to make the predicted cosmic ages consistent with observations.}

\end{quote}


\newpage

\begin{center}
{\em\Large 1. Introduction}
\end{center}

The cosmic age problem \cite{chaboyer} has been a longstanding issue.
For instance,
to accommodate with
two old galaxies 53W091 at $z=1.55$ \cite{Dunlop1}
and 53W069 at $z=1.43$ \cite{Dunlop2},
the cosmological constant $\Lambda $ was employed \cite{Lima1},
which at the same time can be the candidate
for the dark energy \cite{Weinberg1}
driving the currently accelerating expansion of Universe
\cite{Riess1,spergel1,Tegmark1}.
However, the recent discovery of
an old high-$z$ quasar APM 08279 + 5255
with an age $(2.0\sim 3.0)$ Gyr at redshift $z = 3.91$ \cite{Hasinger}
declares the come-back of the cosmic age problem.
With Hubble parameter $h=0.72$ and fractional matter density $\Omega_{m} =0.27$,
the $\Lambda$CDM model would give an age $t=1.6$ Gyr at $z = 3.91$,
much lower than the age lower limit $ 2.0$ Gyr.
This so-called ``high-$z$ cosmic age problem'' \cite{Lima2}
have attracted a lot of attention
\cite{Lima2,Jain,Alcaniz,Cunha,Pires,Capozziello,Rahvar,Wei,Cai},
where various DE models are directly tested against this quasar.
So far there is no DE model that can pass this test,
and most of these models perform even poorer
than $\Lambda$CDM model in solving this issue \cite{Lima2,Jain}.
Besides, the existence of the oldest globular cluster M 107
with an age $15 \pm 2$ Gyr \cite{Tammann}
is also a puzzle.
Using $\Lambda$CDM model,
the latest results of WMAP5 show that
the present cosmic age should be $t_{0} = 13.73 \pm 0.12$ Gyr \cite{Komatsu},
which is lower than the central value of the formation time of M 107.

One can look at this issue from a different perspective.
Instead of studying the age constraints on the specific DE Model directly,
one may firstly study in general
what kind of DE models will be helpful to alleviate cosmic age problem.
In seeking DE models performing better than $\Lambda$CDM model
in solving high-$z$ cosmic age problem,
we find that simply introducing the dark energy alone does not remove the problem,
the scaling law of matter component need be modified,
and the interaction between DM and DE need to be taken into account.
As examples, we consider the age problem in two interacting DE models:
our previously proposed coupled YMC DE model \cite{zhang,Wang} and a interacting scalar DE model.
It is found that both these two interacting DE Models
can predict a cosmic age much greater than that of $\Lambda$CDM model at any redshift,
and thus substantially alleviate the cosmic age problem.
So one may say that the interaction between DM and DE
is the crucial factor required to make the predicted cosmic ages consistent with observations.

The organization of this paper is as follows.
In section 2, we briefly introduce the cosmic age problem.
Since the high-$z$ quasar APM 08279 + 5255 is more difficult to accommodate
than the globular cluster M 107,
the age constraints given by this quasar will be mainly discussed.
In seeking DE models performing better than $\Lambda$CDM,
we find that simply introducing dark energy alone does not remove the problem,
and the interaction between DM and DE need to be taken into account.
In section 3, we consider the age problem in coupled YMC model.
It is seen that the coupled YMC model
predicts a cosmic age much greater than that of $\Lambda$CDM model at redshift $z = 3.91$,
and amply accommodates the high-$z$ quasar APM 08279 + 5255.
Besides, the age constraints given by the globular cluster M 107 is also discussed,
and it is found that coupled YMC model can also substantially accommodate this object.
In Section 4, we consider the age problem in a simple interacting scalar DE model,
and find that this interacting scalar model can also substantially alleviate the cosmic age problem.
In Section 5, we compare our results with previous works,
and discuss the reason why introducing interaction
between DM and DE would be helpful to resolve cosmic age problem.
In this work, we assume today's scale factor $a_{0}=1$,
so the radshift $z$ satisfies $z=a^{-1}-1$.
Throughout the paper,
the subscript ``0'' always indicates the present value of the corresponding quantity,
and the unit with $c=\hbar=1$ is used.

\

\begin{center}
{\em\Large 2. The cosmic age in non-interaction DE models}
\end{center}

The age $t(z)$ of a flat Universe at redshift $z$ is given by \cite{Jain}
\be
\label{00}
t(z)
 =\int_z^\infty\frac{d\tilde{z}}{(1+\tilde{z})H(\tilde{z})},
\ee
where H(z) is the Hubble parameter.
It is convenient to introduce a dimensionless age parameter \cite{Jain}
\be
\label{01}
T(z)\equiv H_0 t(z)
 =\int_z^\infty\frac{d\tilde{z}}{(1+\tilde{z})f(\tilde{z})},
\ee
where
\be  \label{02}
f(z)\equiv H(z)/H_{0}= \sqrt{\Omega_{m}\frac{\rho_{m}(z)}{\rho_{m0}}
+(1-\Omega_{m})\frac{\rho_{y}(z)}{\rho_{y0}}}\, ,
\ee
$H_{0}=100\,h$ km $\cdot $s$^{-1}\cdot$ Mpc$^{-1}$,
and $\rho_{m}(z)$, $\rho_{y}(z)$
are energy density of matter and dark energy,
$\rho_{m0}$, $\rho_{y0}$ are their present values,
$\Omega_{m}$
is the present fractional matter density, respectively.
Here the radiation component has been ignored
since its contribution to $T(z)$ at $z<5$ is only $\sim 0.1\%$.
Once a specific DE model is given,
 $f(z)$ will be known,
and so will be  $T(z)$.
For $\Lambda$CDM model, one has
$\frac{\rho_{m}(z)}{\rho_{m0}}=(1+z)^{3}$,
$\frac{\rho_{y}(z)}{\rho_{y0}}=1$, and
\be \label{03}
 T(z) = \int_z^\infty
 \frac{d\tilde{z}}{(1+\tilde{z})\sqrt{\Omega_{m}(1+\tilde{z})^{3}
 +(1-\Omega_{m})}}\, ,
\ee which gives a larger age for a smaller $\Omega_{m}$. WMAP3 gives
$\Omega_{m}=0.268\pm 0.018$ \cite{Spergel2}, WMAP5  gives
$\Omega_{m}=0.258\pm 0.030$ (Mean) \cite{Dunkley}, and SDSS gives
$\Omega_{m}=0.24\pm 0.02$ \cite{Tegmark2}. For $\Lambda$CDM model,
we list the values of $T(z)$ at redshift $z=3.91$ for various
$\Omega_{m}$ in Table 1.

\begin{table}
\caption{The value  $T(3.91)$ depends on $\Omega_{m}$
in $\Lambda$CDM model.}
\begin{center}
\label{standard model}
\begin{tabular}{|c|c|c|c|c|}
  \hline
  $\Omega_{m}$& 0.27 & 0.24 & 0.21 & 0.17 \\
  \hline
  $T(3.91)$ & 0.118 & 0.125 & 0.133 & 0.148 \\
  \hline
\end{tabular}
\end{center}
\end{table}

At any redshift,
the Universe cannot be younger than its constituents.
Since the high-$z$ quasar APM 08279 + 5255 is more difficult to accommodate
than the globular cluster M 107,
the age constraints given by this quasar will be mainly discussed.
For this quasar, one  must have
\be \label{04}
T(3.91)\geq T_{qua}\equiv H_0 \,t_{qua},
\ee
where $t_{qua}$ is the age of the quasar at $z=3.91$,
determined by the chemical evolution.
The exact value of $t_{qua}$ is not fully determined:
by a high ratio Fe/O from the X-ray result
Ref.\cite{Hasinger} gives $t_{qua} = (2.0\sim 3.0)$ ~Gyr at $z=3.91$,
while  Ref.\cite{Lima2} gives  $t_{qua} = 2.1$ Gyr at the same redshift
by a detailed chemidynamical modeling.
The constraint of Eq.(\ref{04})
sensitively depends on the value of the Hubble constant $h$.
Using the HST key project,
Freedman, et. al. \cite{Freedman} give  $h=0.72\pm0.08$,
and Sandage, et. al. \cite{Sandage} advocate a lower $h=0.623\pm 0.063$.
WMAP3 gives $h=0.732\pm0.031$ \cite{Spergel2},
and WMAP5 gives $h=0.701\pm0.013$ \cite{Dunkley}.
Adopting $t_{qua}=2.0$ Gyr, the values of $T_{qua}$
for different $h$ are listed in Table 2.
It is seen that if $h= 0.64$,
$\Lambda$CDM model can not accommodate
this quasar unless $\Omega_{m}\leq0.21$;
if $h= 0.72$, $\Lambda$CDM model can not accommodate
this quasar unless $\Omega_{m}\leq0.17$.
Since WMAP3 \cite{Spergel2},
WMAP5 \cite{Dunkley},
and SDSS \cite{Tegmark2} give
$\Omega_{m}=0.24$, $0.25$, and $0.24$, respectively,
$\Lambda$CDM has difficulty to accommodate the quasar
even with the age lower limit $t_{qua}=2.0$ Gyr.

\begin{table}
\caption{The value $T_{qua}$ depends on $h$ for fixed $t_{qua}=2.0$ Gyr.}
\begin{center}
\label{quasar age}
\begin{tabular}{|c|c|c|c|c|}
  \hline
  $h$ & 0.72 & 0.64 & 0.58 & 0.57 \\
  \hline
  $T_{qua}$ & 0.147 & 0.131 & 0.119 & 0.117 \\
  \hline
\end{tabular}
\end{center}
\end{table}

\begin{figure}
\centerline{\includegraphics[width=8cm]{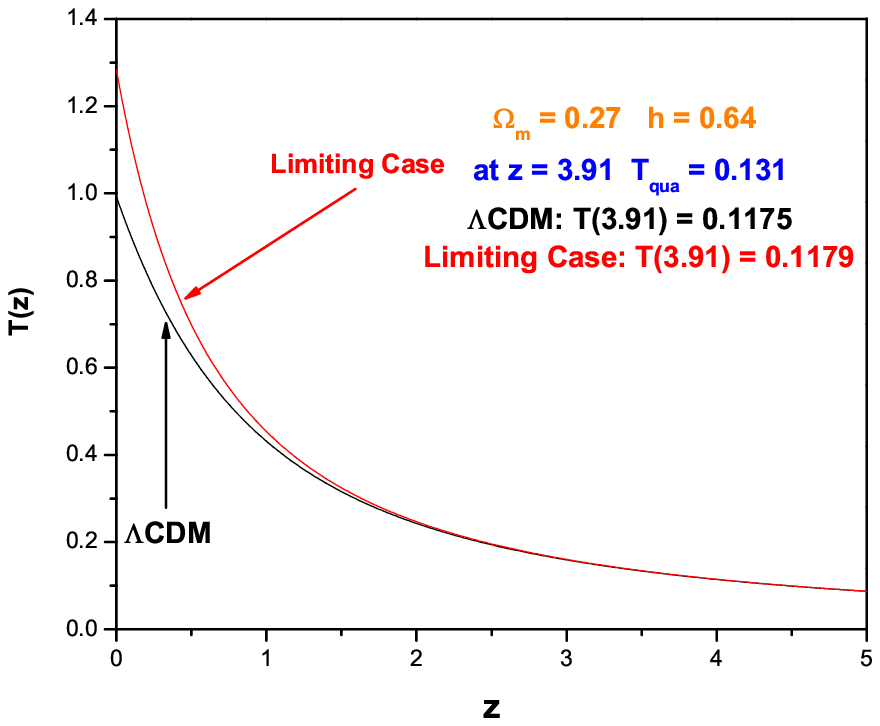}}
\caption{ \label{fig1}
Limiting age  of the Universe in non-interaction  dark energy models.}
\end{figure}

We seek possible DE models, which perform better
than $\Lambda$CDM in solving the age problem.
The key factor is the function $f(z)$ defined in Eq.(\ref{02}),
which should be smaller than that of $\Lambda$CDM.
For the DE models without interaction between DE and matter,
the matter independently
evolves as $\frac{\rho_{m}(z)}{\rho_{m0}}=(1+z)^{3}$.
For the quintessence type DE models,
$\rho_{y}(z)$ is an increasing function of redshift $z$,
then $f(z)$ must be greater than that of $\Lambda$CDM.
Thus this class of models would not work.
On the other hand,
for the phantom type DE models,
$\rho_{y}(z)$ is a decreasing function of redshift $z$,
then a smaller $f(z)$, a larger accelerating expansion,
and an older Universe could be achieved.
However, since the value of $\rho_{y}(z)$ cannot be negative,
the limiting case, as a toy model, is that
$\rho_{y}(z)$ decreases to zero very quickly with increasing $z$.
Then Eq.(\ref{01}) would give a limiting age
$T(z)=\frac{2}{3}(1+z)^{-3/2} \Omega^{-1/2}_{m}$ shown in Fig.\ref{fig1},
where $\Omega_{m}=0.27$ and $h=0.64$ are adopted.
One can see that even this maximal possible age,
which predicted by an extremely phantom type DE model,
is almost the same at $z=3.91$ as that of $\Lambda$CDM.
This is because the DE becomes dominant at very late era,
and thus is not important in calculating the age of early Universe.
Therefore, there is no non-interaction DE model
that can perform better than $\Lambda$CDM.
The main reason is that the matter $\rho_{m}(z)\propto(1+z)^{3}$
growing too fast with redshift $z$ in Eq.(\ref{02}).
To resolve the high-$z$ age problem,
one need to modify the scaling law of matter component;
and the only chance left is
that some DM-DE interaction
might reduce $\rho_m(z)$  and give a smaller $f(z)$.
The following is such two interacting DE models that
substantially alleviate the high-$z$ cosmic age problem.

\

\begin{center}
{\em\Large 3. The cosmic age in Couple YMC Model}
\end{center}

Based on the quantum effective Lagrangian of YM field
\cite{Weinberg2},
the energy density $\rho_y$ and pressure $p_y$ of YMC,
up to the 3-loop, are given \cite{Wang}
\be \label{05}
\rho_y=\frac{1}{2}b \kappa^2e^{y}\left[(y+1)+\eta
(Y_{1}+2Y_{2})-\eta^{2}(Y_{3}-2Y_{4})\right],
\ee
\be
\label{06} p_y=\frac{1}{6}b \kappa^2e^{y}\left[(y-3)+\eta
(Y_{1}-2Y_{2})-\eta^{2}(Y_{3}+2Y_{4})\right],
\ee
where $y=\ln|F/\kappa^2|$,
$F\equiv -\frac{1}{2} F^a{}_{\mu\nu}F^{a}{}^{\mu\nu}= E^{2}-B^{2}$
plays the role of the order parameter of the YMC,
$Y_{1}  =\ln|y-1+\delta|$,
$Y_{2}=\frac{1}{y-1+\delta}$,
$Y_{3}=\frac{\ln^2|y-1+\delta|-\ln|y-1+\delta|}{y-1+\delta}$,
$Y_{4}=\frac{\ln^2|y-1+\delta|-3\ln|y-1+\delta|}{(y-1+\delta)^{2}}$,
$\delta$ is a dimensionless parameter
representing higher order corrections,
$\kappa$ is the renormalization scale with dimension of squared mass,
and for the gauge group $SU(N)$ without fermions,
$b=\frac{11N}{3(4\pi)^2}$, $\eta \simeq 0.8$.
Setting the $\eta^2$ term to zero
gives the 2-loop model,
and setting further the $\eta$ term to zero
gives the 1-loop model.
The cosmic expansion is
determined by the Friedmann equations
\be \label{07}
(\frac{\dot{a}}{a})^2=\frac{8 \pi G}{3}(\rho_y+\rho_m),
\ee
where the above dot denotes the derivative with respect to the cosmic time $t$.
The dynamical evolutions of DE and matter are given by
\be \label{08}
\dot{\rho}_y+3\frac{\dot{a}}{a}(\rho_y+p_y) =-\Gamma \rho_y,
\ee
\be
\label{09} \dot{\rho}_m+3\frac{\dot{a}}{a}\rho_m =\Gamma \rho_y,
\ee
where $\Gamma \geq0$ is the decay rate
of YMC into matter, a parameter of the model.
Notice that the coupled term $\Gamma\rho_y$
reduces the increasing rate of $\rho_m(z)$ with redshift $z$.
To solve the set of equations (\ref{07}) through ({\ref{09}}),
it is convenience to replace the variables ${\rho}_m$, ${\rho}_y$ and $t$
with new variables $x \equiv \rho_m/ \frac{1}{2}b \kappa^2$, $y$, and $N \equiv\ln a(t)$
(see Ref.\cite{Wang} for details).
The initial conditions are chosen at the redshift $z=1000$.
Actually, the age of Universe in Eq.(\ref{01}) is mainly contributed
by the part of $z\leq100$ of integration  with an error   $\sim  1\%$.
We take the parameter $\delta=4$ and the initial $y_{i}=14$.
Given the initial value of $x_i$,
the numerical solution of Eqs.(\ref{07}) through ({\ref{09}}) follows,
and so does $T(z)$ in Eq.(\ref{01}).
To ensure the present fractional densities
$\Omega_{y}=0.73$ and $\Omega_{m}=0.27$,
we take $x_{i}=1.74\times10^{8}$ for a decay rate $\Gamma=0.31 H_0$,
$x_{i}=8.97\times10^{7}$ for $\Gamma=0.67 H_0$,
and $x_{i}=5.28\times10^{7}$ for $\Gamma=0.82 H_0$.

\begin{figure}
\centerline{ \includegraphics[width=8cm]{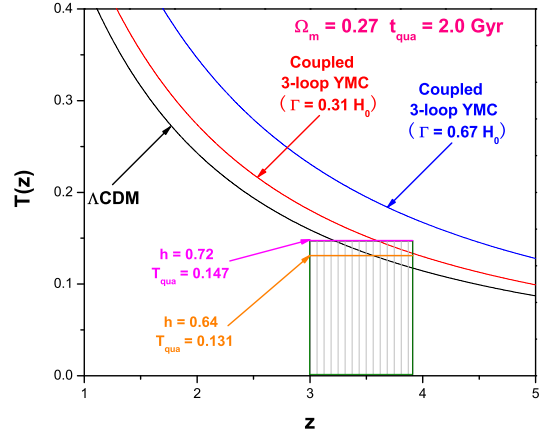}}
\caption{\label{fig2}
The predicted  age $T(z)$ by YMC
and by $\Lambda$CDM.
If $t_{qua}=2.0$ Gyr,
the age problem is amply solved by YMC,
but not by $\Lambda$CDM.}
\end{figure}

We plot $T(z)$ from YMC
and from $\Lambda$CDM in Fig.\ref{fig2},
where $\Omega_{m}=0.27$ and $t_{qua}=2.0$ Gyr are adopted.
The $\Lambda$CDM model
has $T(3.91)=0.118$ and
can not accommodate the quasar for $h\geq 0.58$.
YMC with a large interaction $\Gamma$ will yield a large $T(z)$.
For $\Gamma=0.31H_0$, YMC yields $T(3.91)=0.134$
and accommodates the quasar for $h\leq0.65$.
For a larger interaction $\Gamma=0.67H_0$,
YMC yields $T(3.91)=0.172$
and accommodates the quasar even for $h\leq0.83$.
We also check the 1-loop, and the 2-loop YMC models.
With $\Gamma=0.31H_0$ and $\Omega_{m}=0.27$ fixed,
the 1-loop, and the 2-loop  YMC gives  $T(3.91)=0.128$,
and  $0.133$, respectively;
both perform better than $\Lambda$CDM.
Thus YMC with higher loops of quantum corrections
predicts an older Universe.
In lack of explicit quantum corrections of 4-loops and higher,
$\delta$ has been
used as a parameter to approximately represent that.
A larger $\delta$ will yield a larger $\Omega_{m}$ and a smaller $T(3.91)$.
For instances, taking  $\Gamma=0.31H_0$ and
$\delta=5$ yields $\Omega_{m} \simeq0.31$ and $T(3.91)=0.125$.

\begin{figure}
\centerline{ \includegraphics[width=8cm]{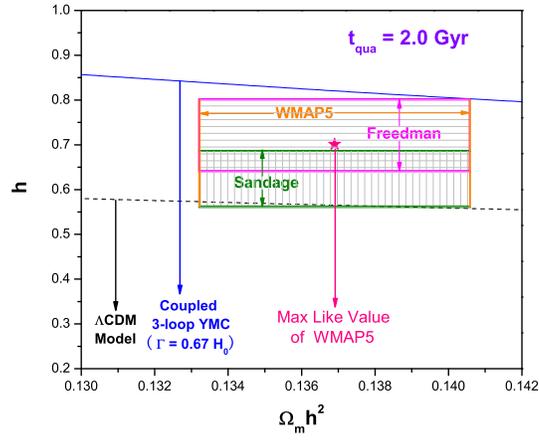}}
\caption{\label{fig3}
The critical curve $h=h(\Omega_{m})$ predicted by YMC and by $\Lambda$CDM
in the $\Omega_{m}h^{2}~-~h$ plane, where $t_{qua}=2.0$ Gyr is adopted.
YMC with $\Gamma=0.67 H_{0}$ amply satisfies the observational constraints.}
\end{figure}

In the flat Universe the age problem is mainly constrained
by the observed data of  $(\Omega_{m},h)$.
Taking a fixed $T_{qua}$ and setting $T(3.91)= \,T_{qua}$,
each specific model has its own critical curve $h=h(\Omega_{m})$.
In Fig.\ref{fig3}, Adopting $t_{qua}=2.0$ Gyr,
we plot the critical curve $h=h(\Omega_{m})$ predicted by YMC and by $\Lambda$CDM
in the $\Omega_{m}h^{2}~-~h$ plane
(Since the WMAP5 constraint on $\Omega_{m}h^{2}$ is much tighter than that on $\Omega_{m}$ alone,
we choose $\Omega_{m}h^{2}$ as the variable of x-axis instead of $\Omega_{m}$),
to directly confront with the current observations.
For each model,
only the area below the curve $h=h(\Omega_{m})$
satisfies $T(3.91)> \,T_{qua}$.
The two rectangles denote the observational constraints:
the first one filled with horizontal lines
is given by WMAP5~+~Freedman \cite{Komatsu,Freedman},
and the second one filled with vertical lines is
given by WMAP5~+~Sandage \cite{Komatsu,Sandage}.
In addition, the star symbol on the plot
denotes the Max Like (ML) value of WMAP5 \cite{Komatsu}.
As is seen, $\Lambda$CDM
can only touch the bottom of WMAP5~+~Sandage,
but is below the ML value of WMAP5 and is out of the scope of WMAP5~+~Freedman.
So it is difficult for $\Lambda$CDM model to accommodate this quasar.
In contrast, YMC model with $\Gamma=0.67 H_0$ passes over all the rectangles,
and amply satisfies the observational constraints.
Therefore, if the quasar has  $t_{qua}=2.0$ Gyr,
the high-$z$ age problem is amply solved in YMC model,
as shown in Fig.\ref{fig2} and Fig.\ref{fig3}.

In our previous works \cite{zhang,Wang},
it is found that
a constant interaction $\Gamma$ corresponds to a constant EOS $w_{0}$,
and a larger $\Gamma$ yields a smaller $w_{0}$.
For instance,
$\Gamma=0.31H_{0}\rightarrow w_{0}=-1.05$;
$\Gamma=0.67H_{0}\rightarrow w_{0}=-1.15$;
and  $\Gamma=0.82H_{0}\rightarrow w_{0}=-1.21$.
In the recent works,
Riess, et. al. advocate an observed constraint $w=-1.02 \pm ^{0.13}_{0.19}$ \cite{Riess07HST}
that gives a lower limit of EOS $w=-1.21$.
In this letter, we shall adopt $w\ge -1.21$,
and this constraint is equivalent to a range $\Gamma=(0\sim 0.82)H_0$ in coupled YMC model.
Taking the upper limit $\Gamma=0.82 H_0$
and adopting WMAP5 ML $\Omega_{m}=0.249$ \cite{Dunkley},
we list the age of the Universe $t_{3.91}$ at $z=3.91$
predicted by YMC for various observed $h$ in Table 3.
YMC accommodates the quasar even with
the age upper limit $t_{qua}=3.0$ Gyr for $h\leq 0.744$;
and even for the upper limit  $h=0.80$ given by Freeman,
YMC still accommodates the quasar if $t_{qua}\leq 2.79$ Gyr.
Moreover, taking the WMAP3 result  $(\Omega_m, h)=(0.238, 0.734)$,
YMC yields the age $t_{3.91} = 3.16$ Gyr at $z=3.91$;
taking the recent WMAP5 ML result $(\Omega_m, h) =( 0.249, 0.724)$,
YMC yields $t_{3.91} = 3.08$ Gyr at $z=3.91$.
For both sets of WMAP data,
YMC with maximal interaction can substantially resolves the high-$z$ cosmic age problem.

\begin{table}
\caption{The age of the Universe $t_{3.91}$ at $z=3.91$ predicted
by YMC with $\Gamma=0.82 H_0$
for various observed $h$ at fixed $\Omega_m=0.249$.}
\begin{center}
\label{YMC limit}
\begin{tabular}{|c|c|c|c|c|}
  \hline
  Observations & Sandage & Freedman & WMAP3 & WMAP5 \\
    \hline
   $h$ & $0.623\pm 0.063$ & $0.72\pm0.08$ &
   $0.732\pm0.031$ & $0.701\pm0.013$ \\
  \hline
  $t_{3.91}$ (Gyr) & $3.25\sim 3.99$  & $2.79\sim 3.49$
       & $2.93\sim 3.18$  & $3.13\sim 3.24$  \\
  \hline
\end{tabular}
\end{center}
\end{table}

Let us turn to the age constraints
given by the globular cluster M 107
with an age $15\pm 2$ Gyr \cite{Tammann},
which is also difficult to accommodate by $\Lambda$CDM model
if the observed data of WMAP are adopted.
To resolve this puzzle,
Sandage and collaborators claim that
a lower Hubble constant $h \sim 0.6$ should be accepted \cite{Tammann,Sandage}.
Although it can solve the problem of M 107,
it still can not solve the high-$z$ cosmic age problem.
Here we list the age of the present Universe $t_{0}$ predicted
by YMC with various interactions $\Gamma$
for the observed data $(\Omega_m, h)$ of WMAP3 and WMAP5 in Table 4.
For the WMAP3 result $(\Omega_m, h)=(0.238, 0.734)$,
YMC with maximal interaction $\Gamma=0.82 H_0$
yields  $t_0=18.5$ Gyr;
for the recent WMAP5 ML result $(\Omega_m, h) =( 0.249, 0.724)$,
YMC with maximal interaction yields $t_0=18.4$ Gyr.
For both sets of WMAP data,
YMC with maximal interaction yields a present cosmic age
that is greater than the age upper limit of M 107,
and thus amply accommodates this oldest globular cluster.

\begin{table}
\caption{The age of the present Universe $t_{0}$ predicted by YMC
with different interaction for the observed data of WMAP3 and WMAP5.}
\begin{center}
\label{YMC current age}
\begin{tabular}{|c|c|c|c|}
  \hline
  Coupled YMC & $\Gamma=0.31 H_0$ & $\Gamma=0.67 H_0$ & $\Gamma=0.82 H_0$ \\
  \hline
  WMAP3 & $14.6$ Gyr & $16.8$ Gyr & $18.5$ Gyr \\
  \hline
  WMAP5 & $14.6$ Gyr & $16.7$ Gyr & $18.4$ Gyr \\
  \hline
\end{tabular}
\end{center}
\end{table}

\

\begin{center}
{\em\Large 4. The cosmic age in a interacting scalar DE model}
\end{center}

As seen above, when the interaction between DM and DE is included,
YMC model can greatly alleviate the cosmic age problem.
One may ask how other interacting DE models will do.
As a specific example, in the following
we shall consider a scalar DE model
with the Lagrangian  \cite{Peebles}
\be \label{A1}
L=\frac{1}{2}\partial^{\mu}\phi\partial_{\mu}\phi-V(\phi).
\ee
We shall consider a simplest case here.
Assuming the potential energy $ V(\phi)$ is dominant,
the energy density and the pressure of DE are
\be \label{A3}
\rho_{\phi}=-p_{\phi} \simeq V.
\ee
The dynamical equations of DE and of matter are
\be \label{A4}
\dot{\rho}_{\phi}=-\Gamma \rho_{\phi},\,\,\,\,\,\,\
\dot{\rho}_m+3\frac{\dot{a}}{a}\rho_m =\Gamma \rho_{\phi}
\ee
where the coupled term $\Gamma\rho_{\phi}$ is introduced.
Since this model is quite similar to the $\Lambda$CDM model,
we name it as coupled $\Lambda$CDM model.
It should be pointed that
this scalar DE model can not resolve the coincidence problem \cite{copeland},
which is naturally solved by the YMC model \cite{zhang,Wang}.
Changing the time variable $t$ to a new variable $N \equiv\ln a(t)$
and making use of the Friedmann equations at $z=0$,
Eq.(\ref{A4}) becomes
\be
\label{A5}
\frac{d\rho_{\phi}}{d N}=-\frac{\sqrt{\rho_{\phi 0}+\rho_{m 0}}}{H_{0}}
\frac{\Gamma \rho_{\phi}}{\sqrt{\rho_{\phi}+\rho_{m}}},\,\,\,\,\,\,\
\frac{d\rho_{m}}{d N} =\frac{\sqrt{\rho_{\phi 0}+\rho_{m 0}}}{H_{0}}
\frac{\Gamma \rho_{\phi}}{\sqrt{\rho_{\phi}+\rho_{m}}}-3\rho_{m}.
\ee
As same as coupled YMC,
the initial conditions are chosen at redshift $z=1000$.
To ensure the present fractional densities
$\Omega_{\phi}=0.73$ and $\Omega_{m}=0.27$,
we take $(\rho_{m i}, \rho_{\phi i})=(1.99\times 10^{8}, 0.76)$ for $\Gamma=0.31 H_0$,
$(\rho_{m i}, \rho_{\phi i})=(1.32\times 10^{8}, 1.52)$ for $\Gamma=0.67 H_0$,
and $(\rho_{m i}, \rho_{\phi i})=(7.12\times 10^{7}, 2.04)$ for $\Gamma=0.82 H_0$.

Here we list the dimensionless age parameter $T(3.91)$ at $z=3.91$ predicted by
coupled YMC and by coupled $\Lambda$CDM model with different interaction in Table 5,
where $\Omega_{m}=0.27$ are adopted.
It is seen that with the same interaction,
coupled $\Lambda$CDM model performs a litter better than coupled YMC model,
and thus also substantially accommodates the old high-$z$ quasar APM 08279 + 5255.
Besides, we also calculate the present cosmic age $t_{0}$ in coupled $\Lambda$CDM model.
adopting $\Omega_{m}=0.27$ and $h=0.72$,
for $\Gamma=0.67 H_0$ and for $\Gamma=0.82 H_0$,
coupled $\Lambda$CDM yields $t_0=16.3$ Gyr and $t_0=18.0$ Gyr, respectively,
both accommodate the oldest globular cluster M 107.
Therefore, one may say that
the interaction between DM and DE
is the crucial factor to make the predicted cosmic ages consistent with observations.

\begin{table}
\caption{The dimensionless age parameter $T(3.91)$ at $z=3.91$
predicted by coupled YMC and by coupled $\Lambda$CDM model with different interaction.}
\begin{center}
\label{more interacting model}
\begin{tabular}{|c|c|c|c|}
  \hline
  Interacting Model & $\Gamma=0.31 H_0$ & $\Gamma=0.67 H_0$ & $\Gamma=0.82 H_0$ \\
  \hline
  Coupled YMC & $0.133$ & $0.172$ & $0.212$ \\
  \hline
  Coupled $\Lambda$CDM & $0.136$ & $0.185$ & $0.246$ \\
  \hline
\end{tabular}
\end{center}
\end{table}

\

\begin{center}
{\em\Large 5. Summary}
\end{center}

We have investigated the cosmic age problem
associated with the old high-$z$ quasar APM 08279 + 5255
and the oldest globular cluster M 107.
To our present knowledge,
there is no interacting DE model performing on this issue had been studied before,
and only non-interaction DE models had been investigated.
For instance,
adopting the age lower limit of APM 08279 + 5255 $t_{qua}=2.0$ Gyr
and a lower Hubble constant $h = 0.64$,
Holographic DE model gives a constraint $\Omega_{m}\leq0.20$ \cite{Wei}
and Agegraphic DE model gives $\Omega_{m}\leq0.22$ \cite{Cai},
both are not better than $\Lambda$CDM model.
It means that in the frame of previous works,
even taking the age lower limit $t_{qua}=2.0$ Gyr,
the sets of WMAP data can not be adopted any more.
In this letter,
it is shown that after introducing the interaction between DM and DE,
a much older Universe can be achieved,
and the data of WMAP can be adopted again
even for the age upper limit $t_{qua}=3.0$ Gyr.
Therefore,
our work greatly improve the previous obtained results.

The introduction of interaction between DM and DE
alleviates cosmic age problem from the following two aspects:
firstly, it yields a EOS $w<-1$,
which correspond to a larger accelerating expansion,
and thus give a larger cosmic age at present;
secondly, it reduces the increasing rate
of matter density $\rho_m(z)$ with redshift $z$,
that would be very helpful to give a smaller $f(z)$ defined in Eq.(\ref{02})
and a larger cosmic age at any redshift.
As seen in Fig.\ref{fig1},
although a larger accelerating expansion would be helpful
to give a larger cosmic age at $z=0$ and solve the problem of M 107,
it still can not solve the problem of high-$z$ quasar APM 08279 + 5255.
This is because the DE becomes dominant at very late era,
and thus is not important in calculating the age of early Universe.
Therefore, the most critical factor that alleviates cosmic age problem is the second one,
i.e. interaction between DM and DE
can modify the scaling law of matter component,
and thus reduce the increasing rate of matter density with redshift $z$.
It is well known that
interacting dark energy models can give
an EOS crossing the phantom divide $w = -1$,
as indicated by the recent preliminary observations \cite{Riess07HST,Astier}.
In this work, we also demonstrate that
the interaction between DM and DE is the crucial factor required
to make the predicted cosmic ages consistent with observations.
Is the DM-DE interaction a necessary factor in describing our real Universe?
This issue still need to be further investigated.

\

ACKNOWLEDGMENT: We are grateful the referee for helpful suggestions.
Y.Zhang's research work is supported by the CNSF No.10773009, SRFDP, and CAS.

\


\newpage

\end{document}